\documentclass[proceedings]{JHEP3} 

\usepackage{epsfig}

\conference{International Workshop on Astroparticle and
High Energy Physics}

\title{Geo-neutrinos, Mantle Circulation and
Silicate Earth}

\author{\speaker{Gianni Fiorentini}\\
Dipartimento di Fisica, Universit\`a di Ferrara, and
Istituto Nazionale di Fisica Nucleare, Sezione di Ferrara,
I-44100 Ferrara, Italy \\
E-mail: \email{fiorenti@fe.infn.it} }
\author{Marcello Lissia\\
Istituto Nazionale di Fisica Nucleare, Sezione di Cagliari, and
Dipartimento di Fisica, Universit\`a di Cagliari,
             I-09042 Monserrato (CA), Italy \\
E-mail: \email{marcello.lissia@ca.infn.it} }
\author{Fabio Mantovani\\
Dipartimento di Scienze della Terra, Universit\`a di Siena,
I-53100 Siena, Italy\\
Centro di GeoTecnologie CGT, I-52027 San Giovanni Valdarno, Italy\\
Istituto Nazionale di Fisica Nucleare, Sezione di Ferrara, 
I-44100 Ferrara, Italy\\
E-mail: \email{mantovani@fe.infn.it} } 
\author{Riccardo Vannucci\\
Dipartimento Scienze della Terra,  Universit\`a di Pavia,
                   I-27100 Pavia, Italy \\
E-mail: \email{vannucci@crystal.unipv.it}
}

\abstract{
In preparation to the experimental results which will be available
in the future, we consider geo-neutrino production in greater
detail than in 
[F.~Mantovani {\it et al.},
arXiv:hep-ph/0309013], 
putting the basis for a more refined model.
We study geo-neutrino production for different models of matter
circulation and composition in the mantle. 
By using global mass balance for the Bulk Silicate Earth,
the predicted flux contribution from distant sources in the 
crust and in the mantle is fixed within 
$\pm 15\%$ (full range). 
A detailed geological and geochemical
investigation of the region near the detector has to be performed,
for reducing the flux uncertainty from fluctuations of the local
abundances to the level of the global geochemical error. A 
five-kton detector operating over four years at a site relatively
far from nuclear power plants can measure the geo-neutrino
signal with 5\% accuracy ($1 \sigma$). 
It will provide a crucial test of the Bulk
Silicate Earth and a direct estimate of the radiogenic contribution
to terrestrial heat.
}
\begin{document}

\section{Introduction}

KamLAND, the Low Energy Anti-Neutrino Detector operating at the
Kamioka mine in Japan, has demonstrated the 
possibility~\cite{Eguchi:2002dm}
of detecting geo-neutrinos, the antineutrinos from decay chains
of radiogenic nuclides inside Earth. A new window on Earth's
interior is thus being opened,
see, \emph{e.g.}, 
Refs.~\cite{Eguchi:2002dm,Fiorentini:2002bp,Fiorentini:2003ww}.

Recently, a reference model of geo-neutrino fluxes has been presented
in Ref.~\cite{Mantovani:2003yd}. 
The model, hereafter indicated as REF, is based on
a detailed description of Earth's crust and mantle, and takes
into account available information on the abundances of Uranium,
Thorium and Potassium --- the most important heat and neutrino
sources --- inside Earth's layers.

This model has to be intended as a starting point, providing
first estimates of expected events at several locations on the
globe, see Fig.~\ref{fig:predflux}
\footnote{An interactive version of the geo-neutrino
event map shown in Fig.~\ref{fig:predflux}
will be available at the site
http://www.fe.infn.it/$\tilde{\mathrm{\ }}$mantovani/geoneutrini/index.htm
.}. 
In preparation to the experimental
results which will be available in the future, from KamLAND as
well as from other detectors which are in preparation, it is
useful to consider geo-neutrino production in greater detail,
putting the basis for a more refined model. In this respect,
let us discuss critically the key ingredients of REF.

\FIGURE{
\epsfig{figure=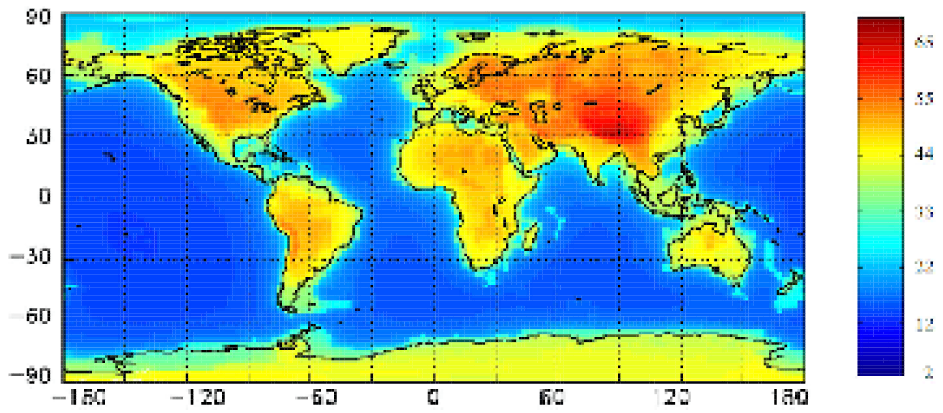,%
width=14cm}
\caption{
  \label{fig:predflux}
Predicted geo-neutrino events form U and Th decay chains
from $\bar{\nu}_{e} + p \to  e^+ + n $ reactions, normalized
to $10^{32}$ proton~yr.
}
}

Concerning the crust, the  $2^{\circ}\times 2^{\circ}$ model of 
Ref.~\cite{mappa} was adopted
for building REF; world-average abundances $a$ of radiogenic
elements have been estimated separately for oceans, the continental
crust (subdivided into upper, middle and lower sub-layers), 
sediments, and oceanic crust.
Although this treatment looks rather detailed on the globe scale,
the typical linear dimension of each tile is of order 200 km,
so that any information on a smaller scale is essentially lost.
Since one expects that about one third of the signal is contributed
from sources within 100 Km from the detector, a better description
of the surrounding region is needed for a refined estimate.

Sandwiched between Earth's crust and metallic core, the
mantle is a 2900 km layer of pressurized rock at high temperature.
As reviewed by Hofmann in Refs.~\cite{hofmann97,hofmann03}, mantle
models can be divided into two broad classes, essentially corresponding
to the presently contradictory geochemical and geophysical evidence
of Earth's interior.

Geochemists have long insisted on a two-layer model, in which
the mantle consists of a relatively primitive layer below a depth
of about 670~km and an upper layer that is highly depleted of heat
producing elements (panel ``a'' in Fig.~\ref{fig:circulation}).
The two layers are viewed
as separate sources of the Mid-Ocean-Ridge Basalts (MORB), which
come from mantle regions that have been already depleted in 
incompatible elements by extraction of the continental crust,
and of Ocean Island Basalts (OIB), which form by melting of
deeper, less depleted or even enriched mantle sectors. 
Also, a more primitive deep layer is needed from
global constraints, otherwise the amount of radiogenic elements
inside Earth is much too small with respect to that estimated
within the Bulk Silicate Earth (BSE) paradigm (see below).

On the other hand, over the past several years seismic tomography
has provided increasingly detailed images of apparently cold
slab descending into the deep mantle, below the 670-km boundary.
If cold slabs descend into the deep mantle, there must be 
a corresponding
upward flow of deep-mantle material to shallow levels (panel
``b'' in Fig.~\ref{fig:circulation}). 
If this circulation reaches the bottom of the mantle
(whole mantle convection), it would destroy any compositional
layering below the crust in a few hundred million years (at a
typical speed of 3~cm~yr$^{-1}$ it takes about 10$^{8}$~yr 
to move down to 2900~km).

\FIGURE{
\epsfig{figure=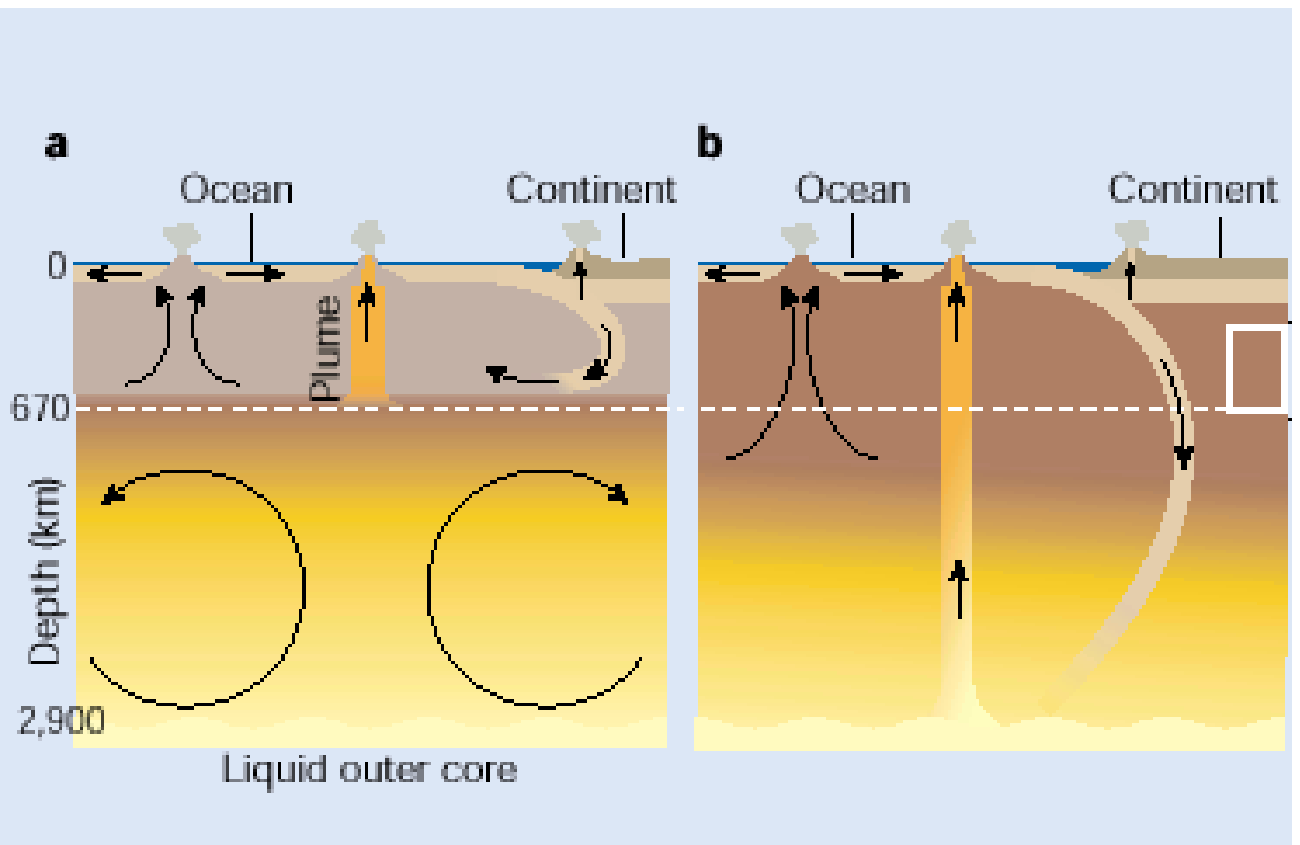,%
width=8cm} 
\caption{
  \label{fig:circulation}
Models of mantle circulation, adapted 
from Ref.~\cite{rama}: 
(a) is the traditional two-layer model with 
demarcation at 670~km and nearly complete isolation 
between upper and lower layers; 
(b) is a fully mixed model.
}
}

In REF, the Preliminary Reference Earth Model~\cite{PREM} was
used for the mantle density profile, dividing Earth's interior
into several spherically symmetrical shells corresponding to seismic
discontinuities. Concerning its composition, a two-layer stratified
model was used: 
for pres\-ent day upper mantle, considered as the source of MORB,
mass abundances of 6.5 and 17.3 ppb for Uranium and Thorium,
respectively, and 78 ppm for Po\-tas\-si\-um  were assumed down to
a depth $h_0=670$~km.  These abundances were obtained by averaging
the results of Refs.~\cite{jochum} and \cite{zartman}.
Abundances below were inferred by
requiring that the BSE constraint is globally satisfied, obtaining
13.2 and 52 ppb for U and Th,  160 ppm for K. In view of the present
debate about mantle circulation and composition, it is desirable
to have a more general treatment, which encompasses both geochemical
and geophysical supported models.

The Bulk Silicate Earth framework 
provides a
compositional description of the primitive mantle, the outer
part of Earth after core separation and before crust differentiation.
In this frame, the amount of heat/neutrino generating material
inside Earth is determined through the following steps:

(a) From the compositional study of selected samples emerging from
the mantle, after correcting for the effects of
partial melting, one establishes the absolute primitive abundances
in major elements with refractory and lithophile character, 
\emph{i.e.},
elements with high condensation temperature (so that they do
not escape in the processes leading to Earth formation) and which
do not enter the metallic core. In this way primitive absolute
abundances of elements such as Al, Ca and Ti are determined,
a factor about 2.75 times CI chondritic abundances.

(b) It is believed, and supported by studies of mantle samples, that
refractory lithophile elements inside Earth are in the same proportion
as in chondritic meteorites. In this way, primitive abundances
of Th and U can be derived by rescaling the chondritic
values~\footnote{Note that Uranium absolute abundances 
in chondritic meteorites are variable within a factor two, 
however the ratio to Aluminum is stable within 10\%.}.

(c) Potassium, being a moderately volatile elements, could have
escaped in the planetesimal formation phase. Its absolute
abundance is best derived from the practically constant 
mass ratio with respect to Uranium observed in
crustal and mantle derived rocks.

The BSE estimates for the total amounts of Uranium, Thorium and
Potassium from different 
authors~\cite{taylor85,hofmann,mcdonough92,wanke84} 
are quite concordant
within 10\%, the central values being 
$m_{\mathrm{BSE}}= 0.8 \times 10^{17}$~kg
for Uranium, $3.1 \times 10^{17}$~kg 
for Thorium, and $0.9\times 10^{21}$~kg for Potassium.
These values can be taken --- within their uncertainties --- 
as representative of the composition of the present crust 
plus mantle system~\footnote{
Geochemists generally agree on the 
absence of these lithophile elements in Earth's core,
see however Refs.~\cite{rama,gessmann}
for a different point of view.}.

Different models can provide different distributions between
crust and mantle, however for each element the sum of the masses
is fixed by the BSE constraint. This clearly provides constraints
on the geo-neutrino 
flux~\footnote{We are always speaking of the flux integrated
over any direction, see~\cite{Fiorentini:2002bp}
for a precise definition of the geo-neutrino flux.} 
which are grounded on sound geochemical
arguments or, alternatively, geo-neutrino detection can provide
a test of an important geochemical paradigm.

We remark that BSE predictions for the radiogenic contributions
to Earth's flow are $H(\mathrm{U})= 7.6$~TW , 
$H(\mathrm{Th}) = 8.5$~TW and 
$H(\mathrm{K}) = 1.8$~TW, 
their sum being about one half of the observed heat
flow from Earth, $H_{E}\approx 40$~TW. Geo-neutrino detection can
thus provide a direct insight on the main source of Earth's 
energetics, see~\cite{Fiorentini:2002bp}.

The reference model was built so as to satisfy the BSE constraint.
However, when discussing its uncertainties a wide range of models
was considered, from a minimal model providing a radiogenic contribution
of just 9 TW up to a fully radiogenic model. These extreme models,
although not excluded from direct observational data, are well
outside the BSE geochemical constraint. If the BSE global constraint
is imposed and only the crust contribution is left free within
the observational uncertainties, then the geo-neutrino prediction
becomes much more precise.

Briefly, in this paper we shall address three questions.

(i) How sensitive are the predicted geo-neutrino fluxes to uncertainties
about the mechanism of mantle circulation?

(ii) Is it possible to test the Bulk Silicate Earth model with
neutrinos?

(iii) Which accuracy is needed concerning the local/regional predictions
and the geo-neutrino detection?

We shall restrict the discussion to geo-neutrinos from Uranium
progeny, which are more easily detectable due to the higher energy.
Extension to the other chains is immediate.

\section{Geochemistry, geophysics and geo-neutrinos}

As mentioned in the introduction, the composition and circulation
inside Earth's mantle is the subject of a strong and so far
unresolved debate between geochemists and geophysicists. Geochemistry
supports the existence of two compositionally distinct reservoirs
in the mantle, the borders between them being usually placed
at depth near $ h_0= 670$~km, whereas geophysics presents evidence
of matter circulation extending well beyond this depth. If this
circulation involves the whole mantle, it would have destroyed
any pre-existing layering, in conflict with geochemical evidence.
In this section we look at the implications of this controversy
on the predicted geo-neutrino fluxes.

One can build a wide class of models, including the extreme geochemical
and geophysical models, in terms of just one free parameter,
the depth $h$ marking the borders between the two hypothetical
reservoirs:

(i) We assume that Uranium abundance in the uppermost part 
of the mantle is nearly chondritic~\footnote{We are
aware that current estimates of U in depleted upper mantle
after crust extraction are in the range of 2 to 
3.9~ppb~\cite{Albarede03,Sun89}. Given the uncertainty
on these values we prefer to perform calculations using
the well-constrained CI value. As shown below, the
assumption of lower U abundance for the uppermost depleted
mantle has limited effects on geo-neutrino flux predictions.
}
($a_{u}= 6.5$ ppb) down to an unspecified depth $h$.

(ii) Below $h$ we determine abundances ($a_{l}$) by requiring
mass balance for the whole Earth.

This means that Uranium mass below the critical depth, 
$m_{>h}$,
is obtained by subtracting from the total BSE estimated mass
($m_{\mathrm{BSE}}$)
the quantity observationally determined in the crust 
($m_{c}$) and that contained in the mantle above $h$
($m_{<h}$):
\begin{equation}
m_{>h} = m_{\mathrm{BSE}} - m_c - m_{<h}
\end{equation}
The abundance in the lower part is then calculated as the ratio
of $m_{>h}$  to Earth's 
mass below $h$  ($M_{>h}$):
\begin{equation}
a_l = m_{>h} /  M_{>h}
\end{equation}

This class of models, described in 
Figs. \ref{fig:tworeservoir} and \ref{fig:Uabunda}, 
includes a fully
mixed mantle, which is obtained for $h = 25$~km (\emph{i.e.}, just below
the crust) so that the strongly impoverished mantle has a negligible
thickness. The traditional geochemical model corresponds to 
$h=h_0$.
As $h$ increases, the depleted region extends deeper inside
the Earth and --- due to mass balance --- the innermost part of the
mantle becomes richer and closer in composition to the primitive
mantle.

\DOUBLEFIGURE{%
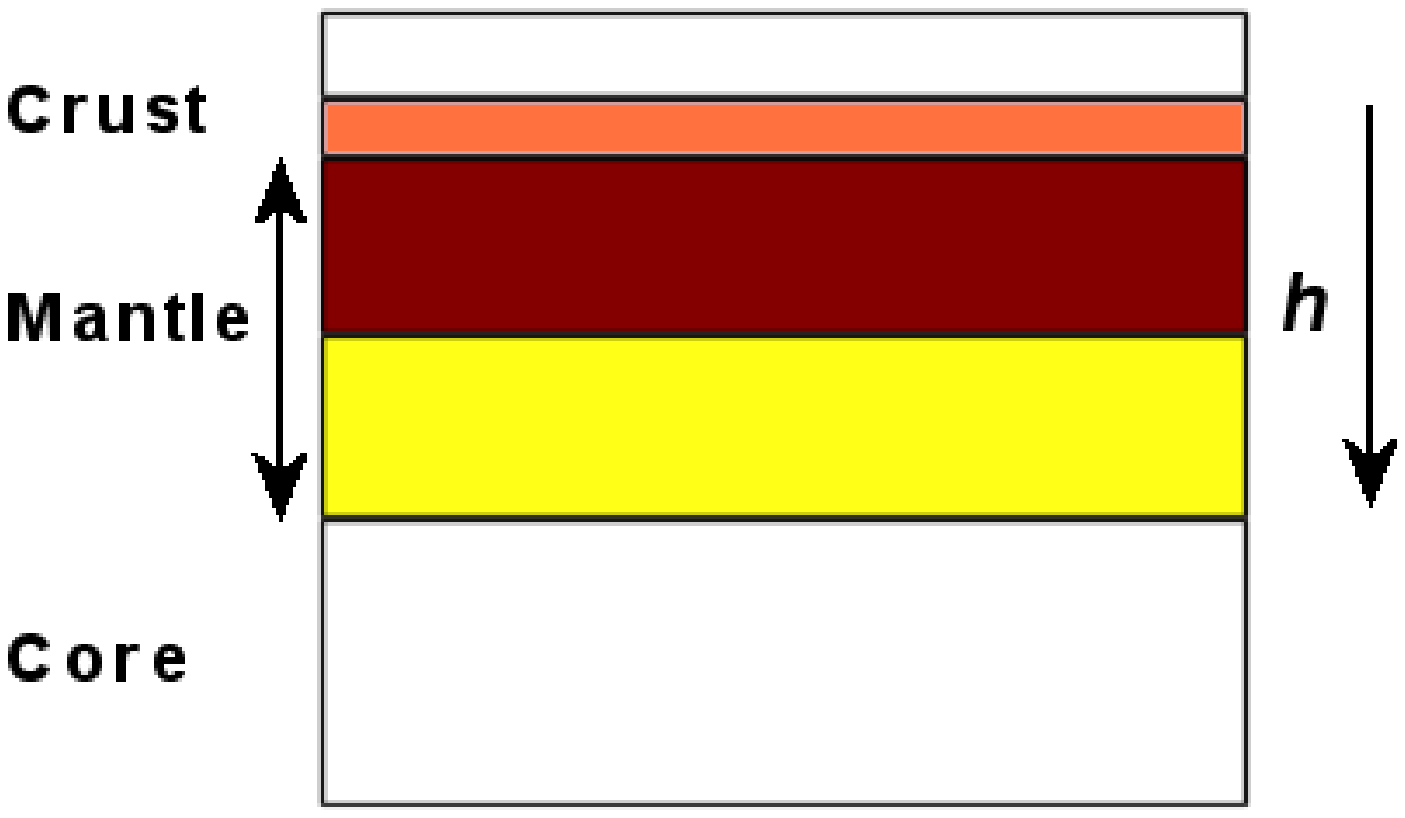,%
width=7cm}  
{%
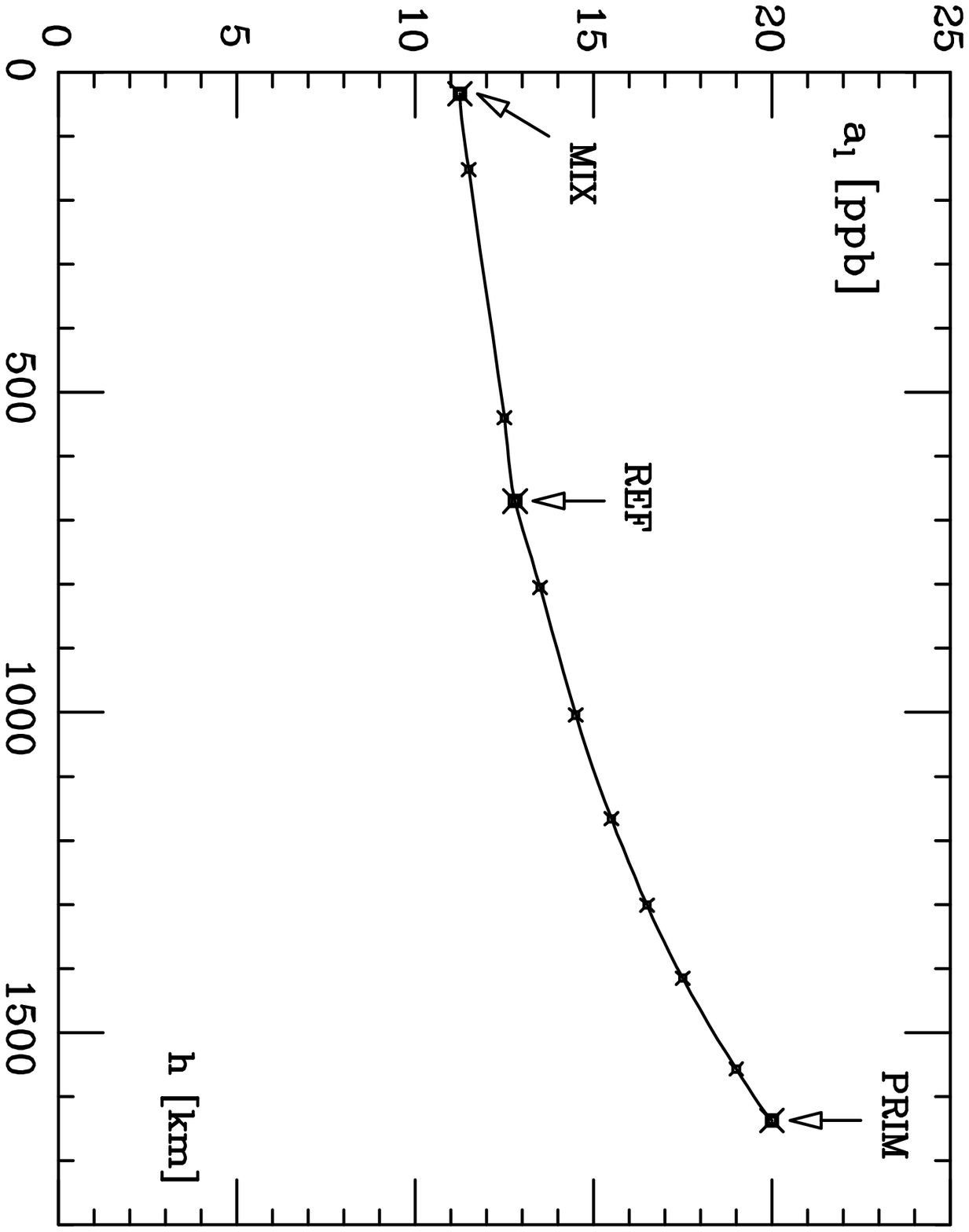,%
width=6.5cm,angle=90}  
{
\label{fig:tworeservoir}
Generic two-reservoir mantle model.
Uranium abundance in the upper part is fixed at
$a_{u} = 6.5$~ppb, the critical depth $h$ 
is a free
parameter, abundance in the lower part $a_l$ is
determined for a fixed total Uranium mass in the
mantle $m_m = 0.45 \times 10^{17}$~kg.
}
{
  \label{fig:Uabunda}
Uranium abundance in the lower part of the mantle
$a_l$ as function of the critical depth $h$
from Earth's surface.
}

Let us discuss in detail a few cases, remembering that the BSE
estimate for Uranium in the whole Earth is 
$m_{\mathrm{BSE}}= 0.8\times 10^{17}$~kg and that 
the best estimate for 
the amount in the crust is 
$ m_c = 0.35\times 10^{17}$~kg~\cite{Mantovani:2003yd} 
so that Uranium in the mantle
is $m_m=0.45\times 10^{17}$~kg.


(a) In the fully mixed model, this quantity has to be distributed
over the mantle mass $M_m= 4.0\times 10^{24}$~kg, which yields
a uniform mantle abundance $a = 11.25$~ppb. We shall refer
to this model as MIX.

(b) If we keep the estimated abundance in the uppermost part 
($a_u = 6.5$ ppb) down to $h_0$ one has the REF 
model \cite{Mantovani:2003yd}.

(c) Among all possible models, the case $h=1630$~km is particularly
interesting. Below this depth the resulting Uranium abundance
is 20 ppb, corresponding to the BSE estimate. The innermost part
of the mantle is thus primitive in its trace element
composition and the crust
enrichment is obtained at expenses of the mantle content above $h$.
We shall refer to this model as PRIM.

Concerning geo-neutrino fluxes from the mantle, all these
models have the same amount of heat/neutrino sources and
only the geometrical distribution is varied. The largest flux
corresponds to the model with more sources close to the surface,
\emph{i.e.} to the MIX model. On the other hand, the minimal prediction
is obtained when the sources are most concentrated at larger
depth, which corresponds to the PRIM case.

\TABLE{
\caption[aaa]
{\label{tab:mantleflux}
Mantle contribution to the produced Uranium geo-neutrino
flux.
The same Uranium mass in the mantle
$m_m = 0.45\times 10^{17}$~kg
and abundance in the upper layer $a_{u}=6.5$~ppb are 
assumed in each model.
}
\begin{tabular}{lcc}
\hline
Model      &     Critical depth $h$ & Flux \\
           &      [km]              &  [$10^6$~cm$^{-2}$ s$^{-1}$] \\
\hline
\hline
MIX        &     25                 &   1.00 \\
REF        &    670                 &   0.95 \\
PRIM       &   1630                 &   0.92 \\
\hline
\end{tabular}
}

From Table~\ref{tab:mantleflux}, 
the difference between the extreme cases is
8\%, model REF being in between. 
The abundance in the upper reservoir $a_{u}$ can also
be treated as a free parameter. If we use an extremely low
value $a_{u}=2$~ppb~\cite{Albarede03} down to about
1200~km and primitive abundance below, we obtain 
the minimal prediction 
$0.86\times 10^{6}$~cm$^{-2}$ s$^{-1}$.

We conclude this section with the following remarks:
\begin{itemize}
\item
uncertainties on the geometrical distribution of trace elements
in the mantle can change the REF prediction for the mantle by
at most $\pm 8\%$.
\item
A geo-neutrino detector at a site where the contribution from
the mantle is dominant (\emph{i.e.}, far from the continental crust)
can be sensitive to the mantle compositional geometry only if
a percent accuracy is reached.
\item
Since at Kamioka mine or at Gran Sasso the mantle contribution
to the total flux is about one quarter of the 
total~\cite{Mantovani:2003yd}, uncertainties
on the mantle geometry imply an estimated error of about
2\% on the total flux predicted with REF.
\end{itemize}

\section{The Bulk Silicate Earth constraint}

So far we have been considering the effect of different geometrical
distributions of trace elements in the mantle, for fixed amount
of these elements within it. Actually the BSE model can be exploited
to obtain tight constraints on the total flux produced
together from the crust and the mantle. In fact, 
with BSE fixing the total
amount of trace elements inside Earth, geometrical arguments
and observational constraints on the crust composition can be
used in order to find extreme values of the produced fluxes.
As an extension of the previous section, the maximal (minimal)
flux is obtained by placing the sources as close (far) as possible
to Earth's surface, where the detector is located.

Concerning Uranium, the range of BSE concentrations reported
in the literature is between 18 and 23 ppb, corresponding to
a total Uranium mass between 
$m(\mathrm{min})=0.72$ and 
$m(\mathrm{max}) = 0.92$ in units of $10^{17}$~kg.

In the same units, we estimate that Uranium mass in the crust
is between  $m_{c}(\mathrm{min}) = 0.30$ and 
$m_{c}(\mathrm{min}) = 0.41$,
by taking the lowest (highest) concentration reported in
the literature \textit{for each layer}, see Table~II
of~\cite{Mantovani:2003yd}. The
main source of uncertainty is from the abundance in the lower
crust, estimated at 0.20 ppm in Ref.~\cite{rudnick95} 
and at 1.1 ppm in Ref.~\cite{shaw86}
Estimates for the abundance in the upper crust are more concordant,
ranging from 2.2 ppm~\cite{condie93} to 
2.8 ppm~\cite{taylor85}. We remark that, within this appraoch,
the resulting \emph{average} crustal U abundance 
$\langle a_{cc}\rangle$ is in the range 1.3-1.8~ppm, which
encompasses all estimates reported in the 
literature~\cite{Weaver84,rudnick95,wedepohl,shaw86} but for that
of Ref.~\cite{taylor85},  $\langle a_{cc}\rangle = 0.91$~ppm,
see Table~\ref{tab:AveCon}~\footnote{Note that
crust mass and flux ranges found
in this paper are tighter than those quoted in
Ref.~\cite{Mantovani:2003yd},
which used the value $\langle a_{cc}\rangle = 0.91$~ppm from
Ref.~\cite{taylor85} as lower limit.}.

\TABLE{
\caption[bbb]
{\label{tab:AveCon}
Average Uranium abundance in the continental crust.
}
\begin{tabular}{lc}
\hline
Reference      &  $\langle a_{cc}\rangle$ [ppm] \\
\hline
\hline
Taylor \& Mclennan 1985~\cite{taylor85}  & 0.91 \\
Weaver \& Tarney 1984~\cite{Weaver84}   &  1.3 \\
Rudnick \& Fountain 1995~\cite{rudnick95}  & 1.42 \\
Wedepohl 1995~\cite{wedepohl}  & 1.7 \\
Shaw {\it et al.} 1986~\cite{shaw86}  & 1.8 \\
\hline
This work, minimal       &  1.3 \\
This work, reference  & 1.54 \\
This work, maximal       &  1.8 \\
\hline
\end{tabular}
}

The highest flux is obtained by assuming the maximal mass in
the crust and the maximal mass in the mantle, 
$m_{\mathrm{tot}}(\mathrm{max}) - m_{c}(\mathrm{max}) = 0.51 $,
with a uniform distribution inside
the mantle, corresponding to $a = 12.8$~ppb.

On the other hand, the lowest flux corresponds to the minimal
mass in the crust and the minimal mass in the mantle, 
$m_{\mathrm{tot}}(\mathrm{min}) - m_{c}(\mathrm{min}) = 0.42 $,
with a distribution in the mantle similar to that of PRIM, 
\emph{i.e.} a strongly depleted mantle with $a_{u}=2$~ppb 
down to about 1300~km and a primordial composition beneath.

The predicted fluxes are shown in Table~\ref{tab:UfluxBSE} 
for a few locations
of particular interest: the Kamioka mine 
($33^{\circ}$N $85^{\circ}$E) where KamLAND is operational, 
the Gran Sasso laboratory ($42^{\circ}$N $14^{\circ}$E)
where BOREXINO~\cite{borexino} is in preparation, 
the top of Himalaya ($36^{\circ}$N $137^{\circ}$E), 
which receives the maximal contribution from the crust, 
and Hawaii ($20^{\circ}$N $156^{\circ}$E), a location where the
mantle contribution is dominant.

\TABLE{
\caption[ccc]
{\label{tab:UfluxBSE}
{\bf Produced Uranium geo-neutrino fluxes within BSE.}
Minimal and maximal fluxes are shown, together with the Reference
values of Ref.~\cite{Mantovani:2003yd}.
Uranium mass $m$ and heat production rate $H$
within each layer are also presented.
Units for mass, heat flow and flux are $10^{17}$~kg, TW and
$10^{6}$~cm$^{-2}$~s$^{-1}$, respectively.
}
\begin{tabular}{lcccccc}
\hline
  &       &        &  Himalaya  &  Gran Sasso &  Kamioka & Hawaii \\
  &  $m$   &  $H$   &  \multicolumn{4}{c}{$\Phi$}    \\
\hline \hline
Crust MIN & 0.30 & 2.85 & 4.92 & 2.84 & 2.35 & 0.33 \\
Crust REF & 0.35 & 3.35 & 5.71 & 3.27 & 2.73 & 0.37 \\
Crust MAX & 0.41 & 3.86 & 6.55 & 3.74 & 3.13 & 0.42 \\
Mantle MIN& 0.42 & 3.99 & \multicolumn{4}{c}{0.80} \\
Mantle REF& 0.45 & 4.29 & \multicolumn{4}{c}{0.95} \\
Mantle MAX& 0.51 & 4.84 & \multicolumn{4}{c}{1.14} \\
\hline
Total MIN & 0.72 & 6.84 & 5.72 & 3.64 & 3.15 & 1.13 \\
Total REF & 0.80 & 7.64 & 6.66 & 4.22 & 3.68 & 1.32 \\
Total MAX & 0.92 & 8.70 & 7.69 & 4.88 & 4.27 & 1.54 \\
\hline
\end{tabular}
}

At any site the difference between the maximal and the minimal
flux predictions are of about 30\%, the extreme values being
within $\pm15\%$ from the reference model prediction.

All this shows the power of the BSE constraint. If the total
amount of Uranium inside Earth is fixed at 
$m_{\mathrm{BSE}} = (0.8 \pm 0.1)\times 10^{17}$~kg,
then the produced geo-neutrino flux at, \emph{e.g.}, Kamioka is
\begin{equation}
\label{fullrange}
\Phi = (3.7 \pm 0.6) \times 10^{6} \mathrm{cm}^{-2} \mathrm{s}^{-1}
\quad\quad\quad \mathrm{(full\ range)}
\end{equation}
after taking into account \textit{the full range} of 
\textit{global} observational
uncertainties on Uranium abundances in the crust and uncertainties
concerning circulation in the mantle.

For a comparison, we remind that in Ref.~\cite{Mantovani:2003yd}, 
where the Uranium mass was allowed to vary in the range 
(0.4--1.7)$\times 10^{17}$~kg, the whole range uncertainty 
on the produced flux at Kamioka was 
$\pm 1.9 \times 10^{6}$~cm$^{-2}$s$^{-1}$.

We remark that the error quoted in Eq.~(\ref{fullrange}) corresponds to a
full range of the predicted values. If, following a commonly
used rule of thumb, we consider
such a range as a $\pm 3 \sigma$ (99.5\%) confidence level,
we deduce a conventional $1 \sigma$ estimate:
\begin{equation}
\label{onesigma}
\Phi = (3.7 \pm 0.2) \times 10^{6} \mathrm{cm}^{-2} \mathrm{s}^{-1}
\quad\quad\quad  (1 \sigma) \quad .
\end{equation}
\section{The effects of geochemical fluctuations and neutrino oscillations}
The main result of the previous section is that --- neglecting
regional fluctuations --- global mass balance provides a precise
determination of the produced geo-neutrino fluxes. We shall compare
this precision with uncertainties resulting from fluctuations
of the regional geochemical composition, from the available information
on neutrino mixing parameters and from the detector finite size.

(a) Actually the Uranium concentration in the region where the
detector is located can be different from the world average and
local fluctuations of this highly mobile element can be envisaged.
These variations, although negligible for mass balance, can affect
significantly the flux. In other words, geochemical arguments
fix the contribution of distant sources, and a more detailed
geological and geochemical investigation of the region around
the detector is needed, the error quoted in Eq.~(\ref{onesigma}) 
providing a benchmark for the accuracy of the local evaluation.

As an example, it has been estimated that about one third
(one half) of the geo-neutrino
signal is generated within a distance of 100 (500) km from Kamioka,
essentially in the Japanese continental shelf. In REF the world
averaged upper crust Uranium concentration $a_{uc} = 2.5$ ppm
was adopted for Japan. In a recent study of the chemical composition
of Japan upper crust~\cite{Togashi2000} more than
hundred samples, corresponding to 37 geologic groups, have
been analyzed.
The composition is weighted with the frequency in the geological
map and the resulting average abundance is $a_{\mathrm{Jap}}=2.32$~ppm , 
which implies a 7.2\% reduction of the flux from Japanese upper crust
with respect to the estimate in REF. Larger variations occur
when rocks are divided according to age or type, see 
Table~\ref{tab:Uabunda},
and even larger differences are found within each group. All
this calls for a detailed geochemical and geophysical study,
with the goal of reducing the effect of regional fluctuations
to the level of the uncertainty from global geochemical constraints.

\TABLE{
\caption[ddd]
{\label{tab:Uabunda}
Uranium abundance in the upper continental crust of Japan.
Groups correspond to rock's age or type and quoted abundances
for each group are area weighted values, 
from Ref.~\cite{Togashi2000}.
}
\begin{tabular}{lcc}
\hline
Group      &  Area (\%)    &  $a$ (ppm) \\
\hline \hline
Pre-Neogene          &    41.7    &  2.20  \\
Pre-Cretaceous       &    10.5    &  2.11  \\
Neog-Quat. Igneous rocks & 24.1   &  2.12  \\
Paleog-Cret.  Igneous rocks & 14.1   &  3.10  \\
\hline
sedimentary & 39.9 & 2.49 \\
metamorphic & 21.3 & 1.72 \\
igneous     & 38.4 & 2.48 \\
\hline
Global area weighted
average              &    99.6    &  2.32  \\
\hline
\end{tabular}
}

(b) Let us remark that the signal is originated from neutrinos
which maintain the electron flavour in their trip from source
to detector, the effective flux being 
$\Phi_{\mathrm{eff}} = \Phi P_{ee}$, 
where $\Phi$ is the produced flux and 
$P_{ee}  = 1 - 1/2 \sin^{2}(2\theta)$ is the (distance averaged) 
survival probability.
From the analysis of all available solar and reactor neutrino
experiments, one gets 
$\tan^{2}\theta = 0.41 \pm 0.04$ at 
$1 \sigma$~\cite{Bahcall:2003ce},
which implies:
\begin{equation}
\label{roadmap}
P_{ee}  = 0.59 \pm 0.02 \quad\quad\quad  (1 \sigma) \quad .
\end{equation}
The survival probability is presently know with a 3\% accuracy, 
so that uncertainties on the neutrino fate do not mask its sources.

(c) If Uranium geo-neutrinos are detected by means of inverse
beta reaction on free hydrogen nuclei 
($\bar{\nu}_e + p \to e^+ + n $) the event number 
is~\cite{Fiorentini:2003ww}
\begin{equation}
N = 13.2\;  \epsilon \frac{\Phi_{\mathrm{eff}}}{10^{6}\, 
\mathrm{cm}^{-2} \mathrm{s}^{-1}} \; 
\frac{N_{p}}{10^{32}} \; \frac{t}{\mathrm{yr}}
\quad ,
\end{equation}
where $\epsilon$  is the detection efficiency, $N_{p}$ is the number
of free protons in the target and $t$ is the measurement time.
For a produced flux $\Phi = 4\times 10^{6}$~cm$^{-2}$~s$^{-1}$ 
and $\epsilon = 0.8$, one expects 25 events for an exposure of 
$10^{32}$ protons~yr.
Statistical fluctuations will be of order $\sqrt{N}$ if
background can be neglected (this clearly does not hold for Kamioka,
due to the many nearby nuclear power plants). In order to reach
a 5\% accuracy --- comparable to that of the global geochemical
estimate --- one needs an exposure of 
$16\times 10^{32}$~protons~yr, which
corresponds to a five-kton detector operating over four years.

\section{Concluding remarks}
We summarize here the main points of this paper:

(1) uncertainties on the geometrical distribution of trace elements
in the mantle (for a fixed mass within it) 
can change the prediction of the reference 
model~\cite{Mantovani:2003yd}
for the geo-neutrino flux from mantle by at most $\pm 8\%$
(full range), the extreme values corresponding to a fully-mixed
and to a two-layer model, with primordial abundance below about
1300 km.

(2) By using global mass balance for the Bulk Silicate Earth,
the predicted flux contribution originating from distant 
sources in the crust and in the mantle is fixed within $\pm 5\%$ 
($1\sigma$) with respect to the reference model. 

(3) A detailed geological and geochemical investigation of the
region within few hundreds km from the detector has to be performed,
for reducing the flux uncertainty from fluctuations of the local
abundances to the level of the global geochemical error.

(4) A five-kton detector operating over four years at a site
relatively far from nuclear power plants can measure the geo-neutrino
signal with 5\% accuracy.

(5) This will provide a crucial test of the Bulk Silicate Earth 
and a direct estimate of the radiogenic contribution to 
terrestrial heat.

\acknowledgments
We express our gratitude for useful discussions Dr. C. Bonadiman,
L. Carmignani, M. Coltorti, S. Enomoto, K. Inoue, E.~Lisi,
T. Mitsui, B.~Ricci, N.~Sleep, A. Suzuki, and F.~Villante.
G.~F. is grateful to the organizing committee of AHEP 2003
for a very enjoiable conference.
This work was partially supported by MIUR (Ministero dell'Istruzione,
dell'Universit\`a e della Ricerca) under 
MIUR-PRIN-2003 project 
``Theoretical Physics of the Nucleus and the 
Many-Body Systems'' 
and  MIUR-PRIN-2002 project ``Astroparticle Physics''. 

\end{document}